\documentclass[aps,prl,twocolumn,showpacs,amsmath,amssymb,superscriptaddress,longbibliography]{revtex4-1}
\usepackage[english]{babel}
\usepackage{amsmath,amsfonts,bm,graphicx,verbatim,mathrsfs,units,color}
\usepackage[colorlinks=true,citecolor=blue]{hyperref}
\usepackage{color}
\newcommand{\revision}[1]{{{#1}}}

\begin{document}

\title{Experimental \revision{Determination} of Dynamical Lee-Yang Zeros}
\author{Kay Brandner}
\affiliation{Low Temperature Laboratory, Department of Applied Physics, Aalto University, 00076 Aalto, Finland}
\author{Ville F. Maisi}
\affiliation{Low Temperature Laboratory, Department of Applied Physics, Aalto University, 00076 Aalto, Finland}
\affiliation{Center for Quantum Devices, Niels Bohr Institute, University of Copenhagen, Universitetsparken 5, 2100 Copenhagen {\O}, Denmark}
\author{Jukka P. Pekola}
\affiliation{Low Temperature Laboratory, Department of Applied Physics, Aalto University, 00076 Aalto, Finland}
\author{Juan P. Garrahan}
\affiliation{School of Physics and Astronomy, University of Nottingham, Nottingham NG7 2RD, United Kingdom}
\affiliation{Centre for the Mathematics and Theoretical Physics of Quantum Non-Equilibrium Systems, University of Nottingham, Nottingham NG7 2RD, United Kingdom}
\author{Christian Flindt}
\affiliation{Low Temperature Laboratory, Department of Applied Physics, Aalto University, 00076 Aalto, Finland}

\date{\today}

\begin{abstract}
Statistical physics provides the concepts and methods to explain the phase behavior of interacting many-body systems. Investigations of Lee-Yang zeros --- complex singularities of the free energy in systems of finite size --- have led to a unified understanding of equilibrium phase transitions. The ideas of Lee and Yang, however, are not restricted to equilibrium phenomena. Recently, Lee-Yang zeros have been used to characterize non-equilibrium processes such as dynamical phase transitions in quantum systems after a quench or dynamic order-disorder transitions in glasses. Here, we experimentally realize a scheme for  \revision{determining} Lee-Yang zeros in such non-equilibrium settings. We extract the dynamical Lee-Yang zeros of a stochastic process involving Andreev tunneling between a normal-state island and two superconducting leads from measurements of the dynamical activity along a trajectory. From the \revision{short-time behavior} of the Lee-Yang zeros, we predict the large-deviation statistics of the activity which is typically difficult to measure. Our method paves the way for further experiments on the statistical mechanics of many-body systems out of equilibrium.
\end{abstract}


\maketitle

\begin{figure}[h!]
\includegraphics[width=0.9\columnwidth]{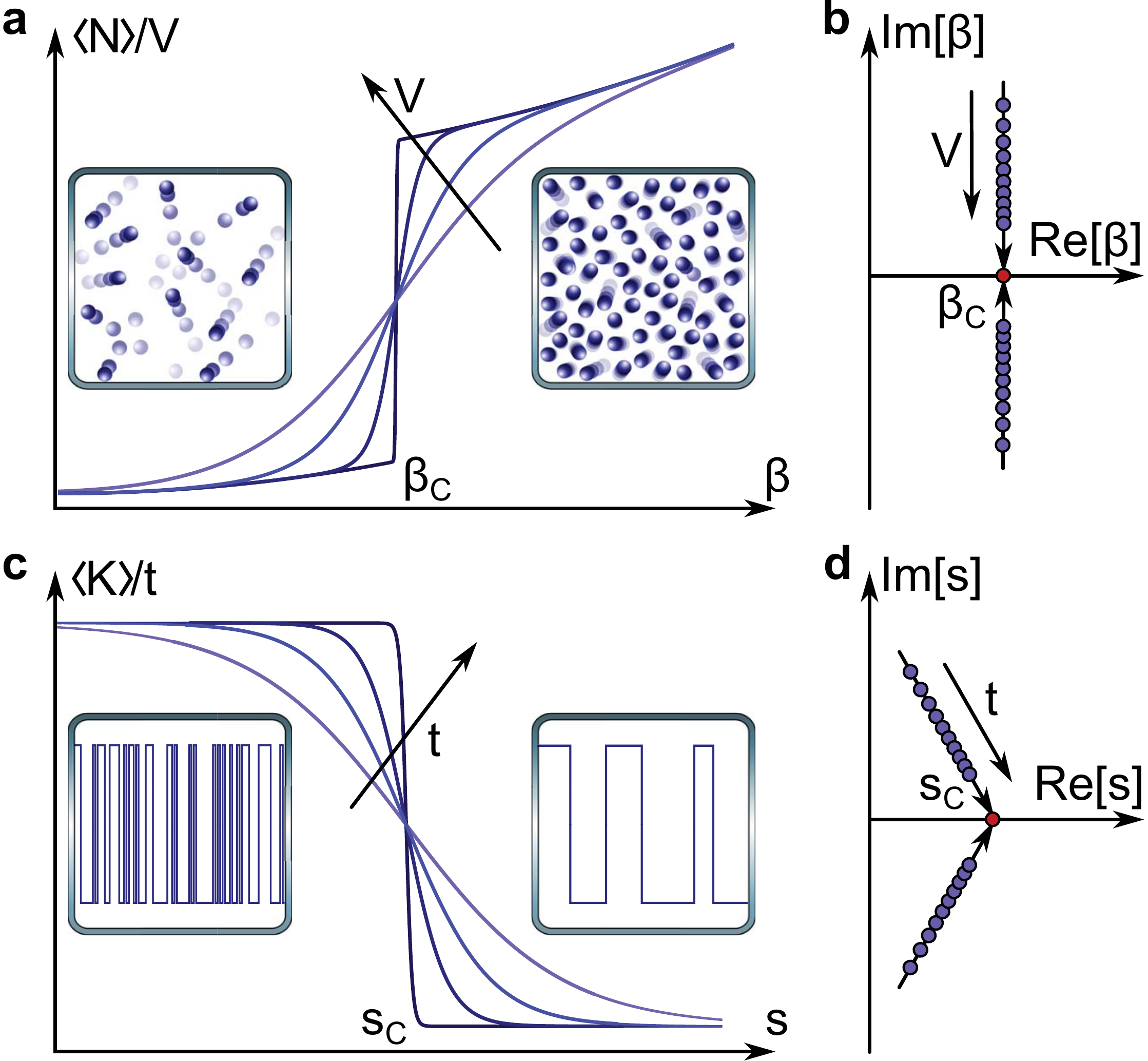}
\caption{Phase transitions and Lee-Yang zeros. \textbf{a,} Density of a gas with $\langle N \rangle$ particles as a function of the inverse temperature $\beta=1/(k_BT)$.  With increasing volume~$V$, the density develops a discontinuity at the critical inverse temperature $\beta_c$ corresponding to a first-order phase transition. \textbf{b,} As $V$ increases, the zeros of the partition function for complex values of $\beta$  (blue dots) approach the critical value $\beta_{{{\rm c}}}$ (red dot) on the real-axis. \textbf{c,} Average activity of an open quantum system as a function of the biasing field~$s$ that couples to the number of random events $K$ along a trajectory of length~$t$. As $t$ increases, the system may exhibit a phase transition between a dynamical phase with a large activity (left box) and one with a low activity (right box). \textbf{d,} With increasing time, the zeros of the dynamical partition function (blue dots) approach the value $s_{{{\rm c}}}$ (red dot) where the phase transition occurs.
\label{FigPT}}
\end{figure}

\emph{Introduction.---}
Phase transitions are ubiquitous physical phenomena involving abrupt changes of a macroscopic system in response to small variations of an external control parameter \cite{Chandler1987,Goldenfeld1992}. A gas, for example, condenses into a liquid when cooled below a certain temperature and its density suddenly increases, Fig.~\ref{FigPT}a. The phase transition is accompanied by large fluctuations of thermodynamic observables and an anomalous behavior of the free
energy~\cite{Callen1985}. Early on, Lee and Yang realized that these exceptional features can be understood from the complex values of the control parameter for which the partition function of a finite-sized system vanishes \cite{Yang1952a,Lee1952}. In the thermodynamic limit, the complex zeros approach the real value of the control parameter for which a phase transition occurs, Fig.~\ref{FigPT}b. Equilibrium Lee-Yang zeros are not only an important theoretical concept \cite{Blythe2003,Bena2005}. They \revision{can also be detected} as demonstrated in a recent experiment on the two-dimensional Ising model \cite{Wei2012,Peng2015}.

The ideas of Lee and Yang have led to a unified understanding of a broad variety of equilibrium phenomena ranging from
percolation \cite{Arndt2001,Dammer2002} and complex networks \cite{Krasnytska2015,Krasnytska2016} to protein folding \cite{Lee2013a,Lee2013b} and Bose-Einstein condensation \cite{Borrmann2000,Dijk2014}. Moreover, it has been recognized that Lee-Yang zeros are not restricted to equilibrium phase transitions. They can also characterize non-equilibrium processes such as dynamical phase transitions occurring in quantum systems after a quench \cite{Heyl2013,Flaschner2016,Azimi2016} or dynamic order-disorder transitions expected in glasses \cite{Merolle2005,Garrahan2007,Hedges2009,Speck2012}.  The partition function is then replaced by a non-equilibrium counterpart \cite{Arndt2000,Blythe2002} and the phase transition may be driven by fields that bias the dynamical trajectories \cite{Merolle2005,Garrahan2007,Hedges2009,Speck2012}, Fig.~\ref{FigPT}c. In general, it can be hard to relate these fields to experimentally controllable parameters. Recently, however, it has been suggested that such phase transitions may be analyzed and detected within the framework of Lee-Yang zeros \cite{Flindt2013,Hickey2014}, Fig.~\ref{FigPT}d.

\begin{figure}
\centering
  \includegraphics[width=0.9\columnwidth]{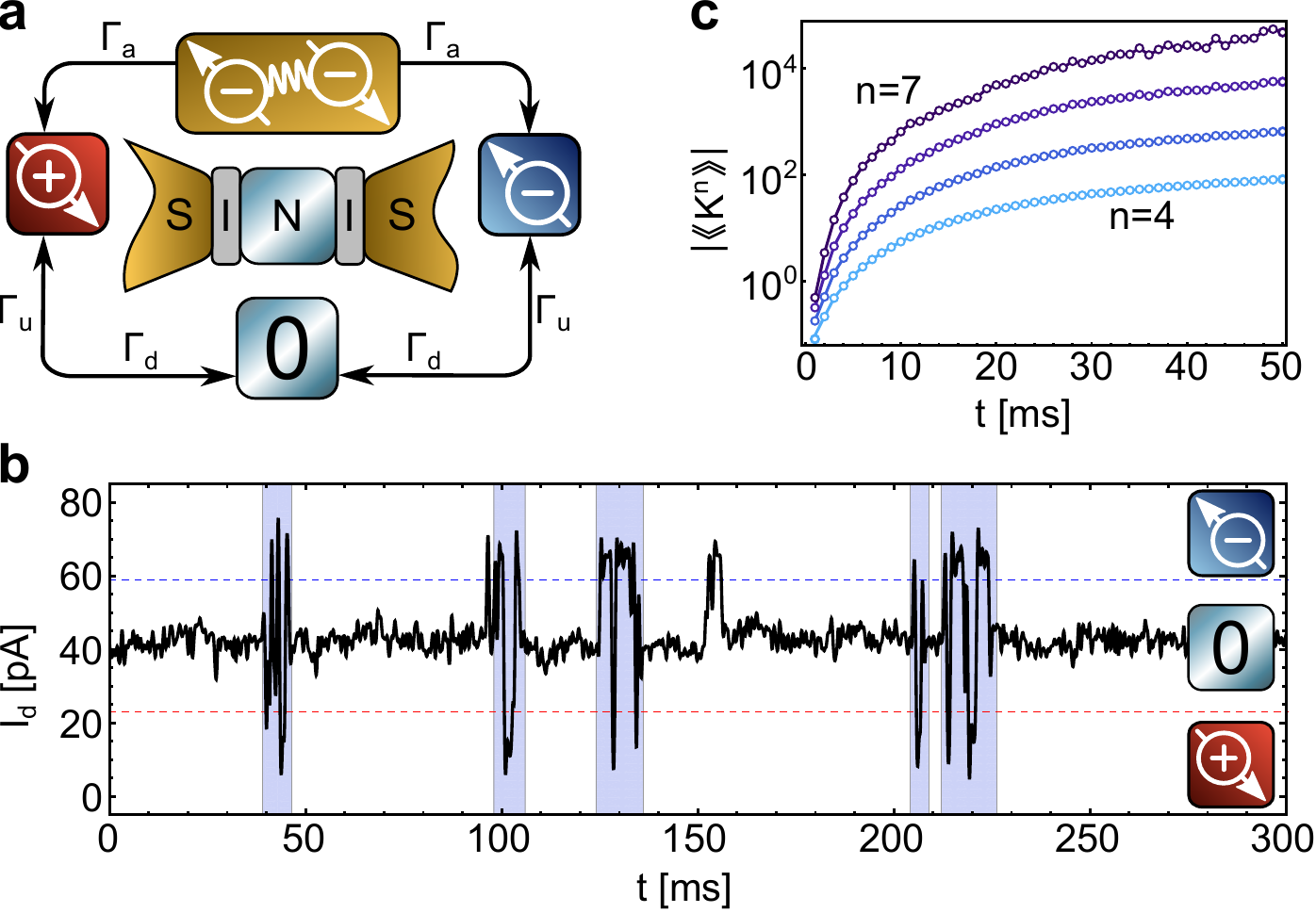}
\caption{Experimental scheme. \textbf{a}, The setup consists of a normal-state metallic island~(N) coupled to superconducting leads~(S) via insulating tunnel barriers~(I) \cite{Maisi2014}. Single-electron events that bring the system from the ground state (0) to one of the excited states ($\pm$) occur with the rate $\Gamma_u=12$~Hz. The reverse process happens with the rate $\Gamma_d=252$~Hz. Andreev events between the excited states happen with the rate $\Gamma_a=615$ Hz. \textbf{b}, The current in a nearby single-electron transistor switches between three levels corresponding to the three charge states of the island. Colored regions indicate Andreev events, where the current switches between a high level (above dashed blue line)  and a low level (below dashed red line).
\textbf{c}, Absolute value of four cumulants $\langle\!\langle K^n\rangle\!\rangle(t)$ of the number of Andreev events along trajectories of length $t$.
\label{FigSystem}}
\end{figure}

In this Letter we experimentally \revision{determine} the dynamical Lee-Yang zeros of a stochastic process in which individual Andreev events are detected and counted in real-time~\cite{Maisi2014}. The charge detection method that we use is by now well-established and it allows for highly accurate statistical measurements \revision{\cite{Gustavsson2006,Gustavsson2007,Gustavsson2009}}.  In each Andreev event, a Cooper pair from a superconductor is transformed into two electrons in a normal-state metal or vice versa. The experimental setup consists of a metallic island coupled to superconducting leads via insulating tunnel barriers, Fig.~\ref{FigSystem}a.  The experiment generates a large ensemble of dynamical trajectories that can be analyzed with the tools of statistical mechanics. Each trajectory is characterized by its dynamical activity \cite{Garrahan2010}, i.~e.~the total number of Andreev events that have occurred, Fig.~\ref{FigSystem}b.  The system switches randomly between two distinct dynamical phases, an active phase with many Andreev events \revision{(transitions between the two excited states)} and an inactive phase with no events \revision{(the ground state)}. We extract the dynamical Lee-Yang zeros from the fluctuations of the activity, Fig.~\ref{FigSystem}c. Importantly, from the \revision{motion} of the dynamical Lee-Yang zeros at short times, we are able to infer their positions in the long-time limit. As we shall see, the Lee-Yang zeros converge to points that are slightly off from the real-axis, corresponding to a smeared phase transition. From the Lee-Yang zeros we can predict the large-deviation statistics of the activity which is otherwise difficult to measure. \revision{Our method can be applied to any experimental setup, where discrete events can be counted and a dynamical partition sum exists \revision{\cite{Gustavsson2006,Gustavsson2007,Gustavsson2009}}.}

\emph{Statistical physics of trajectories.---}
Statistical mechanics describes the equilibrium state of a macroscopic system as an ensemble of microstates realized with a certain probability~\cite{Chandler1987,Goldenfeld1992}.
Likewise, the evolution of a stochastic process can be considered as an ensemble of trajectories characterized by the dynamical partition sum~\cite{Ruelle1999,Gaspard1999,Lecomte2005,Garrahan2007,Esposito2009b,Garrahan2010}
\begin{equation}\label{EqMGF}
Z(s,t)= \sum_K P(K,t)e^{-sK}.
\end{equation}
Here, $P(K,t)$ is the probability of realizing a trajectory of length $t$ with $K$ random events. The intensive field $s$ couples to the extensive observable $K$, similarly to how the magnetic field and the total magnetization are conjugate variables in equilibrium statistical mechanics. The corresponding dynamical free energy is defined as \cite{Garrahan2010}
\begin{equation}\label{EqCGF}
\mathcal{F}(s,t)= \ln Z(s,t).
\end{equation}
For long observation times, the average number of events $\langle K\rangle$ becomes macroscopically large, and a system may exhibit a transition between different phases at a critical biasing field for which the dynamical free energy $\mathcal{F}(s,t)$ becomes non-analytic. Experimentally, however, it is not clear how to tune the biasing field and the measurements presented here are all carried out at $s=0$. Still, it is possible to infer the position of the Lee-Yang zeros from the fluctuations of the activity.

\begin{figure*}
\center
  \includegraphics[width=0.95\textwidth]{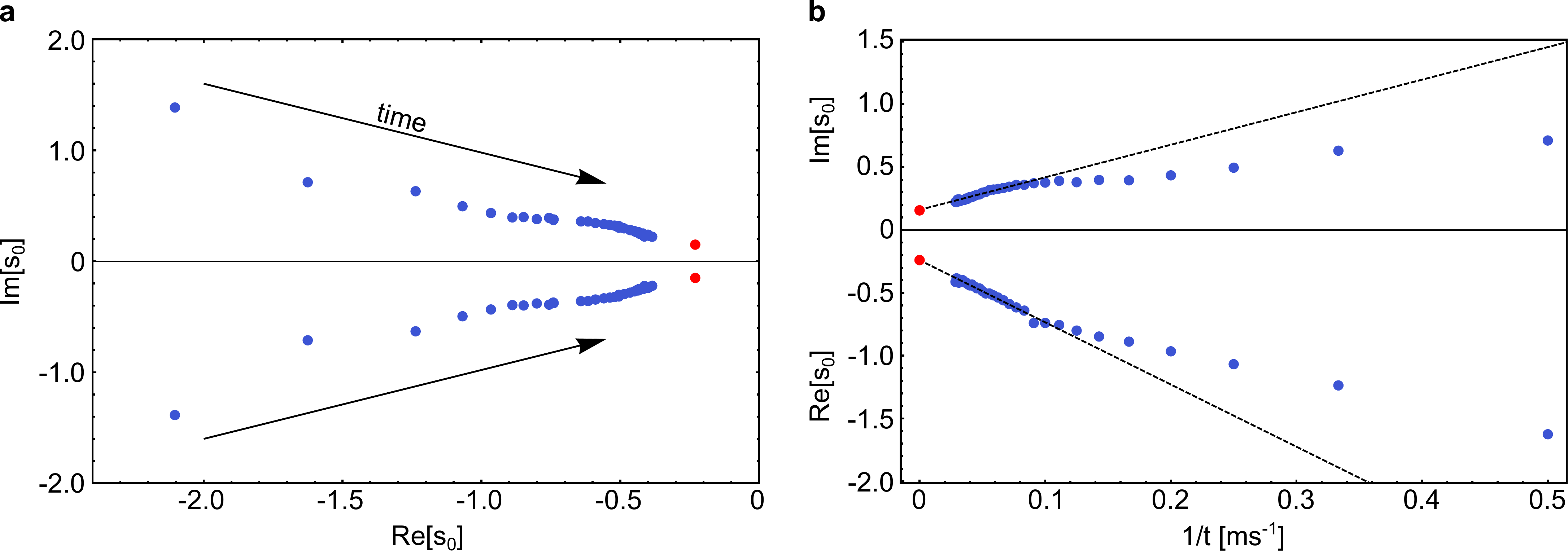}
\caption{\revision{Determination} of dynamical Lee-Yang zeros. \textbf{a,} The leading pair of dynamical Lee-Yang zeros (blue) in the complex plane of the biasing field $s$. The Lee-Yang zeros are extracted from the cumulants of the activity in Fig.~\ref{FigSystem}c for $t=1\, {{{\rm ms}}},\dots,\, 35\,{{{\rm ms}}}$. \revision{The extracted Lee-Yang zeros are unchanged if the cumulant order is increased, ensuring us that sub-leading zeros can safely be neglected. At longer times, sub-leading zeros start to interfere and the extraction method is no longer accurate.} The Lee-Yang zeros converge towards the points $s_c$ and $s_c^*$ indicated in red. \textbf{b,} The convergence points are obtained by linearly extrapolating the dependence of the real part and the imaginary part on the inverse observation time for $t= 10 - 35$ ms \revision{\cite{Flindt2013}}. By taking the inverse time to zero, we find the convergence points $s_{{{\rm c}}}^{(*)} = -0.23\pm 0.15 i$ shown with red in both panels.
\label{FigLYZPlot}}
\end{figure*}

To this end, we consider the cumulants of the activity
\begin{equation}\label{EqCum}
\langle\!\langle K^n\rangle\!\rangle (t)
= \left.(-1)^n\partial^n_s\mathcal{F}(s,t)\right|_{s=0},
\end{equation}
given by the derivatives of the dynamical free energy at $s=0$. In addition, we factorize the dynamical partition sum in terms of its Lee-Yang zeros $s_j(t)$ as \cite{Arfken2001}
\begin{equation}
\label{EqMGFProdExp}
Z(s,t)=e^{sc(t)}\prod_j[s_j(t)-s]/s_j(t).
\end{equation}
Here, the function $c(t)$ depends only on the observation time $t$ and the dynamical Lee-Yang zeros come in complex conjugate pairs since $[Z(s,t)]^\ast=Z(s^\ast,t)$. For $n>1$, the cumulants can now be written as~\cite{Dingle1973,Berry2005,Flindt2009,Flindt2013}
\begin{equation}\label{EqLYZ}
\langle\!\langle K^n\rangle\!\rangle (t) = (-1)^{(n-1)}\sum_j
\frac{(n-1)!}{s_j^n(t)}
\end{equation}
This relation expresses the measurable cumulants in terms of the dynamical Lee-Yang zeros. The high-order cumulants are governed by the pair of Lee-Yang zeros $s_0(t)$ and $s_0^*(t)$ closest to $s=0$, since this pair dominates the sum for large $n\gg1$, such that
\begin{equation}
\langle\!\langle K^n\rangle\!\rangle (t)\simeq (-1)^{n-1}(n-1)!
\frac{2\cos\left[n\;{{\rm arg}}\{s_0(t)\}\right]}{|s_0(t)|^n}.
\end{equation}
From this relation, it follows that the Lee-Yang zeros can be obtained from the expression \cite{Flindt2013}
\begin{equation}\label{EqInversion}
\left[\!\begin{array}{c}
-2 {{\rm Re}}[s_0]\\ |s_0|^2
\end{array}\!\right]=
\left[\begin{array}{ll}
1 & -\frac{\kappa_n^+}{n}\\
1 & -\frac{\kappa_{n+1}^+}{n+1}
\end{array}\!\right]^{-1}
\left[\!\begin{array}{c}
(n-1)\kappa^-_n\\
n\kappa^-_{n+1}
\end{array}\!\right]
\end{equation}
involving the ratios $\kappa^\pm_n(t)= \langle\!\langle K^{n\pm 1} \rangle\!\rangle (t)/\langle\!\langle K^{n}\rangle\!\rangle (t)$ of four successive cumulants. The method is essentially independent of the system under consideration and it can be applied to a variety of equilibrium and non-equilibrium situations. For example, by measuring the cumulants of the magnetization in a spin lattice, one may extract the leading pair of Lee-Yang zeros in the complex plane of the magnetic field.

\emph{Determination of dynamical Lee-Yang zeros.---}
Our experimental setup is shown schematically in Fig.~\ref{FigSystem}a. In the ground state, the normal-state island is occupied by~$0$ (excess) electrons. The tunneling of a single electron between the island and the leads can bring the island to one of its excited states with $\pm 1$ electron. The excited states are energetically degenerate and transitions between them may occur through an Andreev event in which two electrons on the island are transformed into a Cooper pair in one of the leads or vice versa. The charge state of the island is detected using a nearby single-electron transistor whose conductance depends on the number of electrons on the island. By monitoring the current in the single-electron transistor in real-time, we may thus count the individual Andreev events, Fig.~\ref{FigSystem}b. The probability distribution $P(K,t)$ for the number of Andreev events $K$ was measured as a function of the observation time $t$ in Ref.~\onlinecite{Maisi2014}. The corresponding high-order cumulants $\langle\!\langle K^n\rangle\!\rangle (t)$ of order $n=4,5,6,7$ are shown in Fig.~\ref{FigSystem}c. From these four cumulants, we can extract the leading pair of dynamical Lee-Yang zeros using Eq.~(\ref{EqInversion}).

Figure~\ref{FigLYZPlot}a shows the motion of the leading pair of dynamical Lee-Yang zeros in the complex plane of the biasing field $s$. The dynamical Lee-Yang zeros initially move fast, but eventually slow down as they approach the points marked with red. To pinpoint the exact convergence points, we analyze in Fig.~\ref{FigLYZPlot}b the real part and the imaginary part of the dynamical Lee-Yang zeros as functions of the inverse observation time. After an initial transient, \revision{where the Lee-Yang zeros are still far from the convergence points and not yet well-separated}, both the real part and the imaginary part become linearly dependent on the inverse observation time \cite{Flindt2013}. We may then extrapolate the long-time behavior from a linear fit of the experimental data. Specifically, by considering the limit of the inverse observation time going to zero, we can infer the real part and the imaginary part of the dynamical Lee-Yang zeros in the long-time limit. These are the convergence points indicated with red circles in Fig.~\ref{FigLYZPlot}. The convergence points are slightly off the real-axis. The small imaginary part translates to a smeared transition as we discuss below.

\emph{Large-deviation statistics.---}
The extracted convergence points have important implications for the large-deviation statistics of the dynamical activity. To see this, we consider a generic model of a system with two distinct dynamical phases as described by the matrix equation
\begin{equation}
\label{MasterEq}
\frac{d}{dt}|\mathbf{p}(s,t)\rangle=\mathbb{W}(s)|\mathbf{p}(s,t)\rangle
\end{equation}
with the $s$-dependent rate matrix \cite{Jordan2004,Lambert2015}
\begin{equation}\label{RateMatrix}
\mathbb{W}(s)=\left[\!\begin{array}{cc}
\mathcal{H}_1(s) - \Gamma_1 & \Gamma_2\\
\Gamma_1 & \mathcal{H}_2(s) -\Gamma_2
\end{array}\!\right].
\end{equation}
Here, the vector $|\mathbf{p}(s,t)\rangle=[p_1(s,t), p_2(s,t)]^T$ contains the probabilities of being in either of the two phases for $s=0$, where Eq.~(\ref{MasterEq}) reduces to a standard master equation. The fluctuations in each phase are described by the generators $\mathcal{H}_{1,2}(s)$ and switching between them occurs with the rates $\Gamma_{1,2}$. This description is valid for systems with a clear separation of time scales \cite{Gaveau1999,Bovier2002,Macieszczak2016}. Specifically, the inverse switching rates must be much larger than the correlation time of the fluctuations in each phase.

By formally solving Eq.~(\ref{MasterEq}) we can express the dynamical partition sum as
\begin{equation}\label{MGFTP}
Z(s,t)= \langle \mathbf{1}| e^{\mathbb{W}(s)t}|\mathbf{p}_0\rangle,
\end{equation}
where $\langle\mathbf{1}|= [1,1]$ and $|\mathbf{p}_0\rangle$ contains the stationary probabilities given by $\mathbb{W}(0)|\mathbf{p}_0\rangle=0$. Even without knowing the details of the two phases or the switching rates, we can make general statements about the fluctuations. At long times, the dynamical partition sum acquires the large-deviation form $Z(s,t)\simeq e^{\theta(s)t}$, where $\theta(s)=\max{\{\lambda_j(s)\}}$ is the eigenvalue of the rate matrix with the largest real part. Phase transitions are signaled by singularities in the dynamical free energy $\mathcal{F}(s,t)\simeq \theta(s)t$ at the points $s_c$ and $s_c^*$ where the eigenvalues of the rate matrix cross. This is similar to equilibrium phase transitions occurring at eigenvalue crossings of a transfer matrix.

\begin{figure}
  \includegraphics[width=0.95\columnwidth]{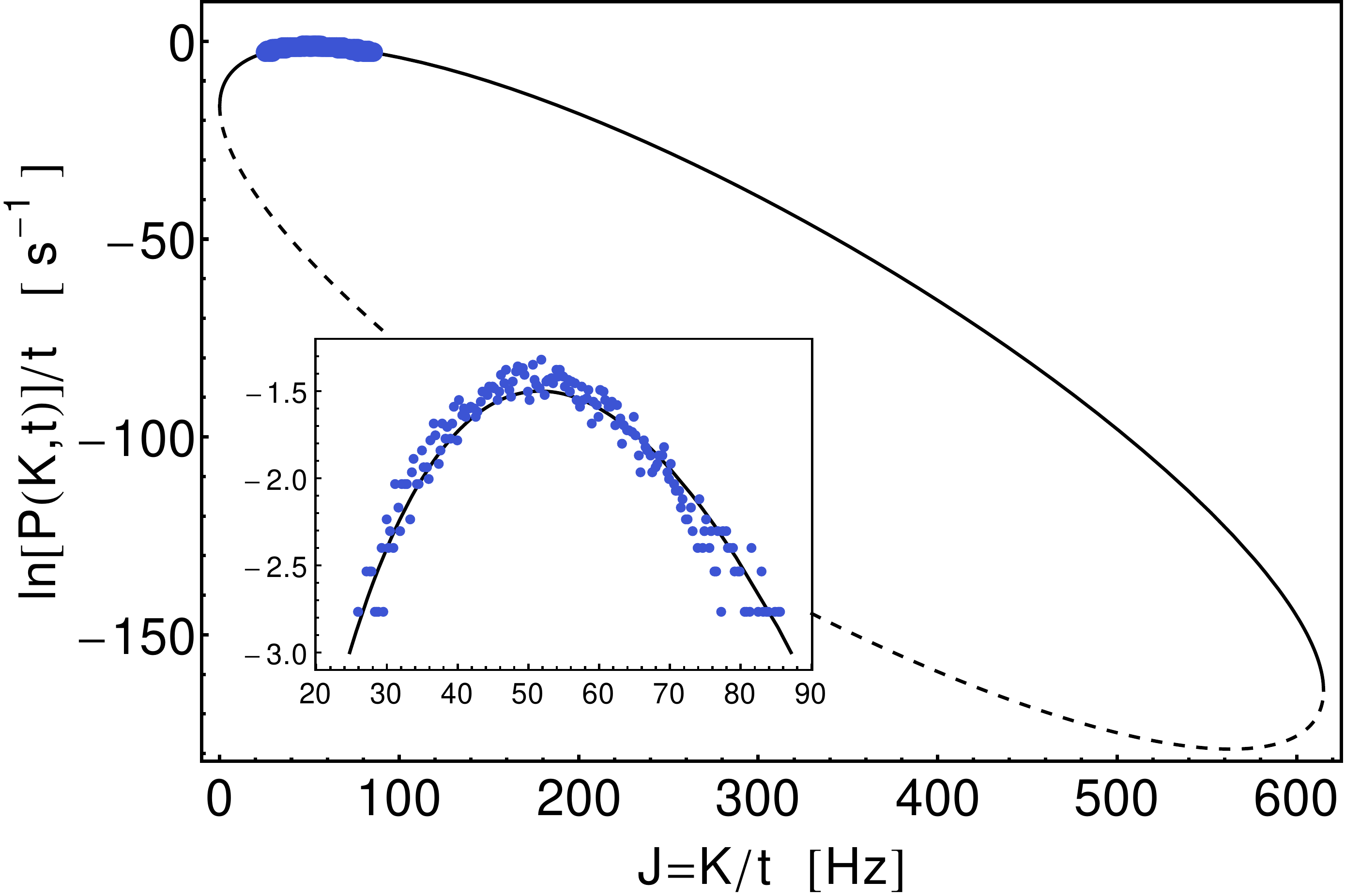}
\caption{Large-deviation statistics of the dynamical activity. The solid line is the upper part of the tilted ellipse in Eq.~\eqref{Ellipse2} using the convergence points extracted in Fig.~\ref{FigLYZPlot}. The dashed line is the lower part of the ellipse. The ellipse is delimited by the average activity in the two phases $J_1\simeq 0$ and $J_2=615$ Hz (the rate at which Andreev events occur between the excited states of the island). To fit the data, the ellipse has been shifted slightly downwards to compensate for an offset which is not included in Eq.~\eqref{Ellipse2}. The distribution measured over $t=3$ s is indicated in blue. \revision{The distribution takes on the large-deviation form at around 3~s which} is much longer than the 35~ms used to extract the convergence points in Fig.~\ref{FigLYZPlot}.
\label{FigLYZPlot2}}
\end{figure}

Additionally, the probability distribution $P(K,t)$ takes the form $\ln[P(K,t)]/t\simeq \varphi(K/t)$ with the rate function $\varphi(J=K/t)$ being related to $\theta(s)$ via a Legendre transformation \cite{Touchette2009}. We now find the general result
\begin{equation}\label{Ellipse2}
\varphi(J)\!=\!-\frac{1}{2}\!
\left(\!\sqrt{\left(|s_{{{\rm c}}}|\!+\!s_{{{\rm c}}}^R\right)|J\!-\!J_2|}
     -\sqrt{\left(|s_{{{\rm c}}}|\!-\!s_{{{\rm c}}}^R\right)|J\!-\!J_1|}
     \right)^2\!\!\!\!,
\end{equation}
having neglected the fluctuations in the individual phases. This is a good approximation when the activity takes values between those that are typical of each phase, i.~e.~for $J_1<J<J_2$, where the fluctuations are mainly due to the random switching. %The tails of the distribution for activities outside this range are determined by the fluctuations in each phase which we disregard as they correspond to faster dynamics which averages out on the time-scales of the interphase switching.
Equation~(\ref{Ellipse2}) is expressed in terms of the convergence points $s_c$ and $s_c^*$ with $s_{{{\rm c}}}^R= {{\rm Re}}[s_{{{\rm c}}}]$. This result is remarkable. It tells us that the convergence points in Fig.~\ref{FigLYZPlot}a, extracted from the short-time observables in Fig.~\ref{FigSystem}c, enable us to predict the large-deviation statistics of the activity. To corroborate this prediction, we show in Fig.~\ref{FigLYZPlot2} the large-deviation function in Eq.~(\ref{Ellipse2}) using the convergence points extracted in Fig.~\ref{FigLYZPlot}. The experiment is in good agreement with the large-deviation function even if it is only feasible to measure the very top of the distribution as fluctuations away from the average are suppressed exponentially in time. The data in Fig.~\ref{FigLYZPlot2} was measured over 3 seconds. This is two orders of magnitude longer than the time needed to extract the convergence points in Fig.~\ref{FigLYZPlot}.% from the behavior of the Lee-Yang zeros.

Geometrically, the shape of the rate function is that of a tilted ellipse~\cite{Jordan2004,Lambert2015} whose width and tilt are determined by $s_c$ and~$s_c^*$. If the convergence points reach the real-axis such that $s_c=s_c^*=s_{{{\rm c}}}^R$ is purely real, the ellipse reduces to the straight line $\varphi(J) = - |s_{{{\rm c}}}^R||J-J_i|$ with $J_i=J_1$ for $s_{{{\rm c}}}^R<0$ and $J_i=J_2$ for $s_{{{\rm c}}}^R>0$.
%This corresponds to the standard Maxwell construction for the probability distribution associated with a first-order phase transition. Specifically, the probability of the activity is that of two dynamical phases, each described by a narrow Gaussian centred at $J_1$ and $J_2$, respectively, connected by an exponential tail due to phase coexistence.
If one could tune the field $s$ across $s_{{{\rm c}}}^R$, there would be an abrupt change in the average activity from $J_1$ to $J_2$, corresponding to a first-order phase transition.  Such a singularity typically only occurs in systems with a large number of degrees of freedom.  (In finite systems, it can occur for trivial reasons, such as when a symmetry splits the dynamics into disconnected ergodic components, see however Ref.~\cite{Nyawo2016}.)  When the convergence points $s_c$ and~$s_c^*$ remain complex, the ellipse has a finite width as in Fig.~\ref{FigLYZPlot2}. In this case, there is a crossover at $s_{{{\rm c}}}^R$ corresponding to a smeared first-order transition. % This is the situation we expect in a system with a finite number of states as the one investigated here.
\revision{By decreasing the switching rates, the transition points would move closer to the origin and a sharp transition would emerge.}

\emph{Conclusions.---}
\revision{We have realized a scheme for determining the leading Lee-Yang zeros in experiment and thereby bridged a gap between theoretical concepts in statistical physics and measurements of fluctuations in many-body systems. Our method can be applied to a large range of equilibrium and non-equilibrium settings, including dynamical phase transitions in quantum systems after a quench or dynamic order-disorder transitions in glasses. As such, our work facilitates several intriguing opportunities for further experiments on the statistical mechanics of many-body systems.}

\emph{Acknowledgements.---}
We acknowledge the provision of facilities by Aalto University at OtaNano Micronova Nanofabrication Centre. Authors at Aalto University were supported by Academy of Finland (project num- bers 284594 and 272218) and are affiliated with Centre for Quantum Engineering. JPG acknowledges support from EPSRC Grant no. EP/K01773X/1.

%\bibliography{library}

%merlin.mbs apsrev4-1.bst 2010-07-25 4.21a (PWD, AO, DPC) hacked
%Control: key (0)
%Control: author (0) dotless jnrlst
%Control: editor formatted (1) identically to author
%Control: production of article title (0) allowed
%Control: page (1) range
%Control: year (0) verbatim
%Control: production of eprint (0) enabled
%

\end{document}